\documentstyle[12pt]{article}                              
\def\word#1{\,\,\mbox{#1}\,\,}
\def\reff#1{(\ref{#1})}
\def\mass{e^{-\phi}{\nabla}^{2}\phi}
\def\nab{{\hat{\nabla}}^{2}\rho}
\def\beq{\begin{equation}}
\def\eeq#1{\label{#1}\end{equation}}

\def\rmunu{R_{\mu\nu}}
\def\dfrac#1#2{{\displaystyle\frac{#1}{#2}}}
\def\del{\delta^{2}(\vec{r})}
\begin{document}
\begin{center}
{\bf Localized Mass and Spin in 2+1 Dimensional\\ Topologically  Massive
Gravity }
\end{center}
\smallskip
\centerline{A. Edery and M. B. Paranjape}
\smallskip
\begin{center}
{\it{Groupe de Physique des Particules, D\'epartement de Physique, 
 \\Universit\'e de Montr\'eal,
C.P. 6128,\\ succ. centreville, Montr\'eal, Qu\'ebec, Canada, H3C 3J7}}
\end{center}
\smallskip
\centerline{Abstract}
\smallskip
{\small Stationary solutions to the full non-linear topologically massive
 gravity (TMG) are obtained for localized sources of mass  $m$ and
 spin $\sigma$. Our results show that the topological term induces spin 
and that the total spin J ( which is the spin observed by an 
asymptotic observer) ranges from $0$ to $\sigma +  
\frac{m}{\mu}\left(\frac{4\pi+m}{4\pi +2m}\right)$ depending on the structure 
of the spin source (here $\mu$ is the topological mass). We find that it is
inconsistent to consider actual delta function mass and spin sources. In the
point-like limit, however, we find no condition constraining $m$ and $\sigma$
contrary to a previous analysis \cite{Clement}.  }

\smallskip
\centerline{I.\quad  Introduction}
\smallskip

A few years ago Deser \cite{Deser} obtained solutions to linearized TMG  with point mass $m$ and point spin $\sigma$ as source.
Among other things, he found that the topological term induced a spin
$m/{\mu}$. In our work we consider the full non-linear theory with a highly
localized (approaching a point) mass source $m$ and 
 spin
source $\sigma$. Like the linear theory, we  obtain an induced spin due to the
presence of the topological term but it has a range of values ( the value
depends on the structure of the spin source $\sigma$). For the case where the
total spin is at its maximum value, we obtain an induced spin of $
\frac{m}{\mu}\left(\frac{4\pi+m}{4\pi +2m}\right)$. As  $m$ gets smaller 
this result
approaches $m/{\mu}$ which is the value obtained in the linear theory.   

Cl\'ement \cite{Clement} has also obtained solutions to the full non-linear
 theory with delta function mass $m$ and delta function spin $\sigma$ as sources. His results showed
that the mass and spin must be constrained by the condition
$m+\mu\sigma =0$. We argue, however, that the use of
a  delta function spin source, while allowed in the linear
theory \cite{Deser}, is not allowed in the full theory and renders the
condition $m+\mu\sigma =0$ invalid.

An alternative approach exists \cite{Gerbert} using dreibeins and the spin
connection, where delta function spin and mass sources can be treated directly.
This formulation facilitates the inclusion of torsion and one finds non-zero
torsion localized at the position of the source.  We will not pursue the
possibility of allowing for torsion below.  Existence of torsion is an
additional structure on the manifold; one has to specify the equation
governing the torsion field.   We restrict our work to metrical theories of
gravitation, where the torsion is zero and the metric is covariantly
conserved.  
  
\smallskip
\centerline{II.\quad Field Equations}
\smallskip

We begin our work by writing down the well known field equations for TMG  with 
energy-momentum tensor $T_{\mu\nu}$ as source.
The Einstein field equations including a topological mass term is given by \cite{Deser,Clement}
(in units where $8 \pi G = 1$) 
\beq
\rmunu \,- \dfrac{1}{2} g_{\mu\nu}R +\dfrac{1}{\mu}C_{\mu\nu} =-{\kappa^2}\, T_
{\mu\nu}
\eeq{ein} 
where $\rmunu$
is the Ricci tensor , $R\,\equiv\,R_{\mu\nu}g^{\mu\nu}$ is the curvature scalar,
and $C_{\mu\nu}$ is the 3 dimensional Weyl (Cotton) tensor defined by
\[
C_{\mu\nu} =\dfrac{1}{2}{\left( detg_{\delta\sigma}\right)}^{{-1}/
{2}}\left(\epsilon_\mu^{\,\alpha\beta}D_\alpha R_{\nu\beta} + \epsilon_\nu^{\,\alpha\beta}
D_\alpha R_{\mu\beta}\right).
\]
Note that the trace of $C_{\mu\nu}$ is identically zero. The reason for the negative sign in front of $T_{\mu\nu}$ in \reff{ein} has been discussed by
Deser \cite{Deser}. We simplify
the field equations \reff{ein} by choosing a rotationally symmetric, stationary metric. The
most general form for such a metric is given by \cite{Vuorio1}
\begin{eqnarray}
ds^2 &=& n^{2} {\left( dt + \omega_{i}\, dx^{i} \right)}^2 + h_{ij}\, dx^{i}
dx^{j}\hspace{0.5cm} i = 1,2 \nonumber\\
&=& n^{2}{\left(\, dt + \psi(r)\,d\theta\,\right) }^2 + h_{ij}\, dx^{i}
dx^{j}
\label{metric}
\end{eqnarray}
with $det\: g_{\mu\nu} = n^{2} \,det\: h_{ij}$. We find that, for the present purpose, it suffices to consider the case with
$n=1$ (i.e. we show that the metric with $n=1$ does support solutions with point mass and point spin).  
In two space dimensions any metric is conformally flat so that we can express $h_{ij}$ as
\beq
h_{ij}= - e^{\phi(r)}\delta_{ij}\,.
\eeq{hij}
The negative sign in \reff{hij} corresponds to Minkowski signature.  The functions $\psi(r)$ and $\phi(r)$ completely determine the metric. We are  particularly interested in the asymptotic behaviour of $\psi(r)$ since it is proportional to the total spin J (see
 \cite{Weinberg}).

The scalar twist $\rho(r)$  is defined by \cite{Vuorio1},
\begin{eqnarray}
\rho & \equiv & \dfrac{n}{\sqrt{det\, h_{ij}}}\epsilon^{ij}\partial_{i}
\omega_{j}\nonumber\\
&=& e^{- \phi(r)}\left(\dfrac{\psi^{'}(r)}{r} + 2\pi\,\psi(0)\del\right)
\label{rho}
\end{eqnarray}
where $\psi^{'}(r)\equiv \dfrac{d\psi}{dr}$. Then the Ricci tensor and curvature scalar are given by 
\begin{eqnarray}
R_{00}&=&\dfrac{1}{2}\rho^2 \nonumber\\ R_{0}^{j}&=&\dfrac{\epsilon^{jk}e^{-\phi}}{2}\,\partial_{k}(\rho)\nonumber\\ R^{ij}&=&\dfrac{1}{2}\hat{R}h^{ij}- \dfrac{h^{ij}\rho^2}{2}\nonumber\\
R&=&\hat{R}- \dfrac{\rho^2}{2}
\label{ricci}
\end{eqnarray}
and the Weyl (Cotton) tensor is given by
\begin{eqnarray}
C_{00}&=&\rho^3 - \dfrac{1}{2}\left(\nab + \hat{R}\,\rho\right)\nonumber\\
C_{0}^{j}&=& \dfrac{\epsilon^{jk}e^{-\phi}}{2}\partial_{k}\left(\dfrac{3}{2}\rho^2 - \dfrac{1}{2}\hat{R}\right)\nonumber\\
C^{ij}&=& h^{ij}\left(\dfrac{-3}{8}\rho^3 + \dfrac{1}{2}\nab + \dfrac{1}{4}\rho\,\hat{R}\right) - \dfrac{1}{2}\hat{\nabla}^{i}\hat{\nabla}^{j}\,\rho
\label{cotton}
\end{eqnarray}
All quantities in \reff{ricci} and  \reff{cotton} are defined with respect to the spatial metric $h_{ij}$. Indices $i,j,k$ are raised and lowered by $h_{ij}$,
$\hat{\nabla}$ stands for covariant differentiation with respect to $h_{ij}$ 
and $\hat{R}$ is the two dimensional curvature scalar given by   
\beq
\hat{R}=\mass \word{where}\nabla^{2}\word{is the flat laplacian.}
\eeq{scalar}
We now solve the field equations \reff{ein} for each component using \reff{ricci}, \reff{cotton} and \reff{scalar}. Without loss of generality we set $\kappa=1$. 

The (0,0), (0,j) and (i,j)
component equations are respectively
\beq
\dfrac{3}{4}\rho^2 - \dfrac{1}{2}\mass + \dfrac{1}{\mu}\rho^3 - \dfrac{1}
{2\mu}\nab - \dfrac{1}{2\mu}\,\rho\,\mass = - T_{00}
\eeq{t00}
\beq
\dfrac{\epsilon^{jk}e^{-\phi}}{2}\partial_{k}\left(\rho + \dfrac{3}{2\mu}
\rho^2 - \dfrac{1}{2\mu}\mass\right) = -  T_{0}^{j}
\eeq{t0j}
\beq
-e^{-\phi}\delta^{ij}\left(-\dfrac{\rho^2}{4}- \dfrac{1}{2\mu}\rho^3 +
 \dfrac{1}{2\mu}\nab + \dfrac{1}{4\mu}\,\rho\,\mass\right)- \dfrac{1}{2\mu}\hat{\nabla}^{i}\hat{\nabla}^{j}\,\rho = - T^{ij}\,.
\eeq{tij}

Note that the scalar twist $\rho$ appearing in the above equations cannot contain a delta function or else quantities like $\rho^2$ and $\rho^3$ in 
those equations would be ill defined. This implies that $\psi(0)$ appearing in the definition of $\rho$ i.e. \reff{rho} must be zero.
Therefore $\rho$ reduces to
\beq
\rho = e^{-\phi}\,\dfrac{\psi^{'} (r)}{r} .
\eeq{rho2}

We approach the problem of localized sources, not by actually specifying
$T_{\mu\nu}$ 
but by examining the metric dependent side of the field equations
(\ref{t00}-\ref{tij}), and drawing conclusions on the scalar twist $\rho$ and
the function $\phi$ if $T_{\mu\nu}$ were localized. Hence $T_{\mu\nu}$ is only
defined via the field equations (\ref{t00}-\ref{tij}), and therefore
automatically covariantly conserved (i.e. since the metric dependent side of the field equations obey the Bianchi identities).  

\smallskip
\centerline{III.\quad Solving the Field Equations}
\smallskip

We consider a localized spin source $\sigma$ by allowing the scalar twist 
$\rho(r)$ be a rapidly decreasing function of $r$.                                                  
The total spin J, which is conserved and
invariant under general coordinate transformations, is given by (see
\cite{Weinberg} )
\beq
J \equiv  2\pi \, \lim_{r \rightarrow \infty} \psi(r)= 
2\pi \int_0^\infty \rho e^{\phi} r dr \, ,
\eeq{total}
where  \reff{rho2} was used. The scalar twist $\rho$ can therefore be interpreted as the total spin density.    
The mass $m$ is defined as the total energy when the total spin  
is zero (i.e. $\rho =0$) and when the topological term is absent 
(i.e. $\mu \rightarrow \infty$). It is 
therefore given by
\beq
m = \int \dfrac{1}{2} \left( \mass \right) e^{\phi} \, d^2 r =
\int \dfrac{1}{2}\, \nabla^2 \phi \, d^2 r \,.
\eeq{m}
Here $\nabla^2 \phi$ can be arbitrarily
localized. The above equation for $m$ leads to the condition that
\beq
\left. r\,\phi^{'}\,\right|_0^\infty = \dfrac{m}{\pi}\quad .
\eeq{limits}
For reasons given in section IV we do not allow  $\nabla^2 \phi$ to be 
a delta function (i.e. we exclude the possibility that $\phi(r)
\propto \ln r$ as $r\rightarrow 0$). Therefore 
\beq
\lim_{r\rightarrow 0}(r\,\phi^{'}) = 0
\eeq{lim0}
i.e. if $\lim_{r\rightarrow 0}(r\,\phi^{'}) = k
\word{where} k\ne 0 , \word{then} \phi(r)\propto k\ln r \word{as} r\rightarrow 
0$.  

We now want to find an expression for the total spin $J$. We first integrate the $T_0^j$ equation \reff{t0j} and find that  
\beq
 \int \epsilon^{ij} \, x^i \, \left( - T_{0}^{j}\right) e^{2\phi} \, d^2 r 
=\dfrac{-1}{2} \int \left( \rho + \dfrac{3}{2\mu} \rho^2 
 -  \dfrac{1}{2\mu}
\,\mass \right)^{\prime} \, e^{\phi}\,r \,d^{2}r \,.
\eeq{sigma}  
The integral of the third term
can be readily evaluated using \reff{limits} and \reff{lim0} and gives 
$-\left(m/\mu\right)\left(1+m/4\pi\right)$. As we will show in section IV, 
this result is different from that obtained using naive manipulations with 
delta function sources. For the first two terms we take  $\rho(r)$ to be a rapidly decreasing function 
of r  with the condition that 
$\lim_{r\to\infty}\rho(r) r^2 e^{\phi} = \lim_{r\to\infty}\rho(r)^2 r^2 
e^{\phi} =0$.
The integrals of these two terms are well defined for any regular mass 
distribution $\nabla^{2}\phi$, but depend on the actual profile and there will
be small corrections for any well localized mass.  However, when evaluated in 
the point mass limit the result is 
\beq
 \left( 2\pi +m \right) \int_0^\infty \left( \rho + \dfrac{3}{2\mu} \rho^2 \right) e^{\phi} r dr\,.
\eeq{parts}
Then using \reff{total} for the total spin, one can rewrite  
\reff{sigma} as   
\beq 
J  = \sigma + \,
\dfrac{m}{\mu}\left(\dfrac{4\pi+m}{4\pi+2m}
\right) - 2\pi \int_0^\infty \dfrac{3}{2\mu} \rho^{2} e^{\phi} r dr \,.
\eeq{J}
where
\beq
\sigma \equiv \dfrac{2\pi}{2\pi+m} \int \epsilon^{ij} \, x^i \, \left( - T_{0}^{j}\right) e^{2\phi} \, d^2 r .
\eeq{sigfig} 
Here $\sigma$ is identified as the spin source (or bare spin) i.e. it is equal to the total spin when the topological term is absent ($\mu\rightarrow\infty$).
Clearly, as  $m$ approaches zero equation \reff{sigfig} reduces to the usual definition of spin in the linear theory
\cite{Deser}. The $m$ appearing in the denominator of the spin source $\sigma$ 
arises because the spatial metric describes a cone with a negative angular
defect
 $m$ and therefore the angle ranges from $0$ to $2\pi +
m$ instead of
$2\pi$ (see \cite{Deser}). 

The last
two terms on the right hand side of  \reff{J} can be regarded as the induced 
spin, which has a 
dependence on $\rho(r)$. The total spin $J$ can therefore range from $0$ to its maximum
value of $ \sigma +  \frac{m}{\mu}\left(\frac{4\pi+m}{4\pi+2m}\right)$. The 
induced spin reduces to 
the linearized result $m/\mu$ in the limit
where $m$ and $\rho$ are small.

We do not obtain any specific equation relating the mass $m$
to the spin $\sigma$ as in ref.\cite{Clement} where the condition
$m+\mu\sigma=0$ is obtained. For the specific 
case $\rho=0$ (which implies $J=0$) we obtain
$m\left(\frac{4\pi+m}{4\pi+2m}\right) +\mu\sigma = 0$ in disagreement
with the result $m+\mu\sigma=0$ of Ortiz \cite{Ortiz}. We now discuss the 
reasons for this discrepency. 
  
\smallskip
\centerline{IV.\quad The Point Like Limit}
\smallskip

We now give an analysis of the use of delta functions for sources. A delta function spin source 
for $T_0^j$ was used in the work of Cl\'ement \cite{Clement}
\beq
T_0^j = \dfrac{-1}{2} \sigma e^{-\phi} \epsilon^{jk} \partial_k 
\left( e^{-\phi} \, \del \,\right)
\eeq{clem}
where $\sigma$ is equivalent to the spin source defined in \reff{sigfig}. 
Then  \reff{t0j} becomes
\beq
\partial_k \left( \rho + \dfrac{3}{2\mu}\rho^2 -
\dfrac{1}{2\mu} \mass \right) = \sigma \partial_k \left( 
e^{-\phi} \, \del \,\right)\,.
\eeq{partial}
The $\rho$ and $\rho^2$ terms on the left hand side of \reff{partial}
cannot match the delta
function on the right hand side (see comment above \reff{rho2}) 
and for the purposes of
our forthcoming argument will simply be dropped).  
The term $\mass$, with $\phi$ proportional to 
$\ln r$, cannot match the delta function because the non-linearity in  
\reff{partial} imposes the products of derivatives of $\ln r$ which are
ill defined at the origin. Consider the contraction of \reff{partial} with 
$x^{k}$.
This implies
\beq
- \dfrac{1}{2\mu}e^{-\phi} \left(x^{k}\partial_k \left( \nabla^{2}\phi\right) 
- r\phi^{'}\nabla^{2}\phi\right)= \sigma e^{-\phi} 
\left(x^{k}\partial_k\del - r\phi^{'}\del\right) 
\eeq{contract}
which requires, for $\phi=-(\mu\sigma/\pi) \ln r$,
\beq
 r\phi^{'}\nabla^{2}\phi =-2\mu\sigma \,\,r\phi^{'}\del 
\eeq{contract2}
where the first term in \reff{contract} is properly matched and the 
$e^{-\phi}$ is canceled with the corresponding  $e^{\phi}$ in any volume 
element.
Equation \reff{contract2} is not sensible for the imposed choice 
$\phi\propto \ln r$. We cannot simply cancel the  $r\phi^{'}$ from
both sides since it is ill defined exactly at the point where the delta
function has all its weight. It is not consistent to take  $r\phi^{'}$ to be
identically constant while taking $\nabla^{2}\phi \equiv (r\phi^{'})^{'}/r$ to
be a delta function. The product  $r\phi^{'}\nabla^{2}\phi$ is clearly ill
defined for $\phi\propto\ln r$. The right hand side of \reff{contract2} is 
also ill defined for $\phi\propto\ln r$ because  $r\phi^{'}$ does not lie in
the space of functions on which $\del$ acts.

A reasonable way to handle the product $r\phi^{'}\nabla^{2}\phi$ is to use the
equivalence                                                      
between a delta function $\del$ and the limit
of a sequence of functions $\delta_{n}(r)$ (see \cite{Arfken}) where
\beq
\int \delta_{n}(r)\,d^{2}r = 1 \,\,\,\forall\, n \,\,\word{and}\,\,
\lim_{n \to \infty}\int \delta_{n}(r)f(r)\,d^{2}r = f(0).
\eeq{delnr}
Here $\delta_{n}(r)$ can be any smooth function, like a gaussian,  that peaks a
$n\to \infty$. 
Then to satisfy \reff{m}                                          
we let
\beq
\nabla^{2}\phi_{n}(r) = 2m \delta_{n}(r)\,.
\eeq{2mdel}
Clearly \reff{limits}
\beq
\left. r\,\phi^{'}_{n}\,\right|_0^\infty = \dfrac{m}{\pi}\quad 
\eeq{limits2}
is valid for each $n$. Since
$\nabla^{2}\phi_{n}(r)$ is a smooth function for every $n$, it follows that
$\phi_{n}(r)$ for any given $n$ cannot be proportional to $\ln r$ as $r\to 0$
(since  
$\nabla^{2}\ln r$ is not smooth at r=0).                              
Hence $\lim_{r\to 0}\,(r\phi^{'}_{n})=0$. 
Then the moment 
\begin{eqnarray}
M &\equiv &\dfrac{-1}{2} \int x^{k} \partial_k \left( \dfrac{-1}{2\mu}\mass 
\right)  e^{\phi}\, d^{2}r\nonumber\\&=& -\dfrac{1}{4\mu}\int 
\left(2 + r\phi^{'}\right)\nabla^{2}\phi \,d^{2}r
\label{moment}
\end{eqnarray}
which was used to calculate the third term of \reff{sigma} yields
\begin{eqnarray}
M &=&
\dfrac{-\pi}{2\mu}\lim_{n\to\infty}\int_0^\infty
\dfrac{1}{r}\left(r\phi^{'}_{n}\right)^{'}\left(2+r\phi^{'}_{n}\right) r \,dr
\nonumber\\&=&\dfrac{-\pi}{2\mu}\lim_{n\to\infty}
\left(2\left.\left(r\phi^{'}_{n}\right)                                                                                                             
\right|_0^\infty + \dfrac{1}{2}\left.\left(r\phi^{'}_{n}\right)^2 \right|_0^\infty
\right)\nonumber\\&=&\dfrac{-\pi}{2\mu}\lim_{n\to\infty}
\left(2\dfrac{m}{\pi}+\dfrac{1}{2}\left(\dfrac{m}{\pi}\right)^{2}\right)
\nonumber\\
&=& -\dfrac{m}{\mu}\left(1+\dfrac{m}{4\pi}\right)
\label{goodcal}
\end{eqnarray}
where the limits on $r\phi^{'}_{n}$ i.e.
\beq                                                    
\lim_{r \to \infty}r\phi^{'}_{n} = \dfrac{m}{\pi}\word{and}
\lim_{r \to 0}r\phi^{'}_{n}= 0
\eeq{rphi}
was used (also note that the result \reff{goodcal}
is valid for any $n$ and is independent 
of the limit $n\to\infty$). The right hand side of \reff{contract2} would be equal to zero if the
functions $\phi_{n}(r)$ were used, showing clearly the impossibility of 
satisfying equation \reff{contract2} (and similarly \reff{partial}). 
A naive attempt
to satisfy this equation by having $\phi = (m /\pi)\ln r$,
$r\phi^{'}= m/\pi$ and
$\nabla^{2} \phi = 2m \del$,
which is not a consistent treatment of $r\phi^{'}$, leads to the
incorrect conclusion that $m +\mu\sigma=0$ (see \cite{Clement}) and to a 
wrong value
for the moment $M$, namely
\beq
M = -\dfrac{m}{\mu}\left(1
+\dfrac{m}{2\pi} \right)\,.
\eeq{wrong}
The factor $\left(\frac{4\pi +m}{4\pi +2m}\right)$ in    \reff{J}
was obtained using the result \reff{goodcal} and would not appear if the result
\reff{wrong} were used. This is why for the specific case $\rho=0$
we obtain the condition $m\left(\frac{4\pi+m}{4\pi+2m}\right) +\mu\sigma = 0$
and not                                                                          
the condition $m+\mu\sigma=0$ of Ortiz \cite{Ortiz}.

We have shown that the correct procedure in the non-linear theory is to take the source to be a function $\delta_{n}(r)$ that peaks as
$n\to\infty$ instead of starting directly with a delta function. In the linearized theory \cite{Deser} such a procedure is not necessary
because products like $ e^{-\phi}\nabla^{2}\phi$ do not appear in the
equations and one can begin directly with a delta function source. These 
products actually reflect the non-linearity of 
the field equations and this is why the
factor $\left(\frac{4\pi +m}{4\pi +2m}\right)$ in \reff{J} 
does not appear in
the linearized theory.   

We thank Nourredine Hambli for useful discussions and NSERC of Canada and FCAR
du Qu\'ebec for financial support.

\end{document}